\documentclass[twocolumn,english,prb,epsfig,psfig]{revtex4}
\usepackage[T1]{fontenc}
\usepackage[latin1]{inputenc}
\usepackage{graphicx}

\makeatletter

\usepackage{babel}
\makeatother
\begin{document}

\title{Negative refraction and superlensing in a 2D photonic crystal structure
}

\author{R. Moussa$^{1}$, S. Foteinopoulou$^{1}$, Lei Zhang$^{2}$, G. Tuttle$^{2}$, K. Guven$^{3}$, E. Ozbay$^{3}$, and C. M. Soukoulis$^{1}$}

\address{$^{1}$Ames Laboratory-USDOE and Department of Physics and Astronomy,Iowa
State University, Ames, Iowa 50011}
\address{$^{2}$Department of Electrical and Computer Engineering and Microelectronics Research Center,Iowa
State University, Ames, Iowa 50011}
\address{$^{3}$Department of Physics, Bilkent University, Bilkent, 06533 Ankara, Turkey}
\begin{abstract}
We experimentally and theoretically studied a new left-handed (LH)
structure based on a photonic crystal (PC) with a negative refractive
index. The structure consists of triangular array of rectangular dielectric bars with dielectric constant $9.61$. Experimental and theoretical results demonstrate the
negative refraction and the superlensing phenomena in the microwave
regime. The results show high transmission for our structure for a wide range of incident angles. Furthermore,
surface termination within a specific cut of the structure excite
 surface waves at the interface between air and PC and allow the
reconstruction of evanescent waves for a better focus and better transmission. The normalized
average field intensity calculated in both the source and image planes
shows almost the same full width at half maximum for the source
and the focused beam.
\end{abstract}
\maketitle
Left-Handed-Materials (LHM) are materials with simultaneously negative
dielectric permittivity $\varepsilon\left(r\right)$ and negative
magnetic permeability $\mu\left(r\right)$. The phase velocity of the
light wave propagating inside this medium is pointed in the opposite
direction of the energy flow. Thus, the Poynting vector and wave vector
are antiparallel, consequently, the light is refracted negatively.
The existence and impact of such materials was first pointed out by
Veselago \cite{key-1}. Years later, several theoretical and experimental
groups investigated the LHM\cite{key-2,key-3,key-4,key-5}. For LHM
based on PCs, Notomi studied light propagation in a strongly modulated
two-dimensional (2D) PC\cite{key-6}. Luo et al have studied subwavelength
imagining in PC\cite{key-7}. Cubukcu et al demonstrated experimentally
single-beam negative refraction and superlensing in the valence band
of 2D PC operating in the micro wave regime\cite{key-8}. Foteinopoulou
et al emphasized the time evolution of an EM wave as it
hits the interface between a right-handed (RH) and a LH material interface\cite{key-9}.
Pendry was the first who suggested that a slab of lossless LHM with
both $\varepsilon\left(r\right)$ and $\mu\left(r\right)$ equal to
-1 should behave as a perfect lens\cite{key-10}, i.e, the small details
as well as the larger ones could be reproduced by such a lens. The
reconstruction of the evanescent wave components is the key to such
perfect focusing. It is shown in another study \cite{key-11} how
the evanescent waves get amplified upon reaching the interface between
a RH and LH medium and consequently how they participate in improving
the quality of the image. Homogeneous metamaterials and PCs were used
to demonstrate the effect. The main challenge is to find a structure
with $n=-1$ for which the matching condition are verified with no
reflections. The candidate structure should have three characteristics.
First, it should exhibit almost isotropic Equal Frequency Surfaces
(EFS) in the band with negative curvature\cite{key-5}. Secondly, it should guarantee
the absence of any higher order Bragg reflections for any angle of
incidence. Finally it should guarantee single beam propagation.

In
this paper we propose such a structure based on PC. The proposed structure
meets all the required conditions and has a refractive index
very close to $n=-1$ as well as a high transmission  at the desired frequency.
Additionally, with an appropriate cut and a specific termination of
the surface, the proposed structure focuses the image in a better
way by exciting the surface waves and enabling them to contribute
toward the quality of the image. Furthermore, for different incident angles, the cut structure demonstrates a high transmission as well as a maximum coupling with the surface modes at the interfaces. The proposed structure consists of a triangular
array of dielectric bars with a dielectric constant $\varepsilon=9.61$
in air. The dimensions of each bar in the x and y directions are respectively
$r_{x}=0.40a$ and $r_{y}=0.80a$ where $a=1.5875$~cm is the center
to center separation between bars. The length of each bar is $l=45.72$~cm.
Fig.~1 shows the structure with more details. Only the transverse magnetic
(TM) modes are considered here ( the $\mathbf{E}$ field is parallel to the rods). 

\begin{figure}[b!]
\includegraphics[%
  scale=0.7]{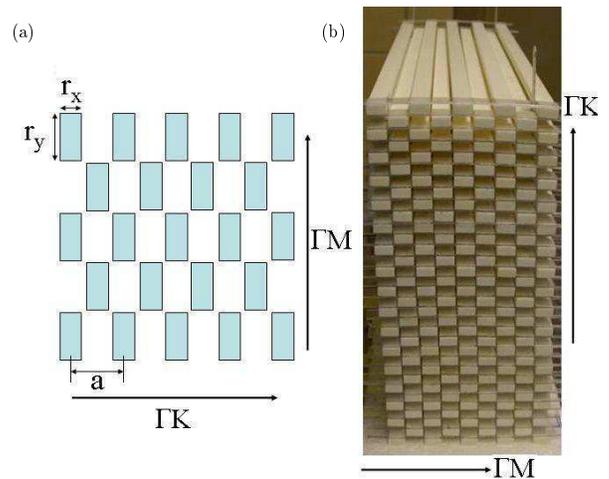}

\caption{(a) A triangular array of dielectric bars in air with $r_{x}=0.40a$
and $r_{y}=0.80a$, where $a=1.5875$~cm is the lattice constant. (b)
A picture of the structure with 33 layers in the lateral direction 
and 9 layers in the propagation direction.}
\end{figure}

The plane wave method is used to compute the photonic band structure
as well as the EFS. Some of the theoretical results are obtained using
the Finite Difference Time Domain (FDTD) method. This method shows
a time and space evolution of the emitted EM waves. More details of
this algorithm can be found in reference \cite{key-12}. In all the
FDTD simulations we report, LIAO boundary conditions\cite{key-13}
are used. Depending on the desired conditions for the case under study,
the source emits a monochromatic TM- (E-) polarization of desired dimensionless
frequency. A Gaussian source (Gaussian in space and \char`\"{}almost\char`\"{}
monochromatic) is placed outside the structure to check the negative
refraction. However, a point source also excites the surface
waves at the interface between RH and LH material. The photonic band
structure as well as the EFS in \textbf{k} space are shown in Fig.~2.
The quantity $(fa/c)$ is the dimensionless frequency where
$c$ is the velocity of light and $a$ is the lattice constant. The second band, extending from $0.27$
to $0.40$, has a negative curvature and is of particular interest.
To study the system and to compute the refractive index, the EFS are
plotted for different frequencies. Note that the EFS consist of the
allowed propagation modes for a specific frequency and in our case, they have closed concave like form. Within the second
band the shape of the EFS tends to became circular for higher frequencies
and reproduces the symmetry of the system for lower frequencies. It is important to notice that in contrary with the triangular lattice with circular cross section for which the hexagonal symmetry is conserved on the lattice  as well as on the scatterers, this structure with rectangular rods lose its hexagonal symmetry for the scatters when the wave is traveling along a specific direction. Thus the breaking of the symmetry gives some anisotropy in the structure. Thus, the EFS are not totally isotropic because of the breaking of the symmetry. Therefore, the computed refractive index of our structure using the EFS plot is somehow affected by the anisotropy and is not exactly equal to $-1$ instead, it ranges from $-0.97$
to $-1.22$.

\begin{figure}
\includegraphics[%
  scale=0.9]{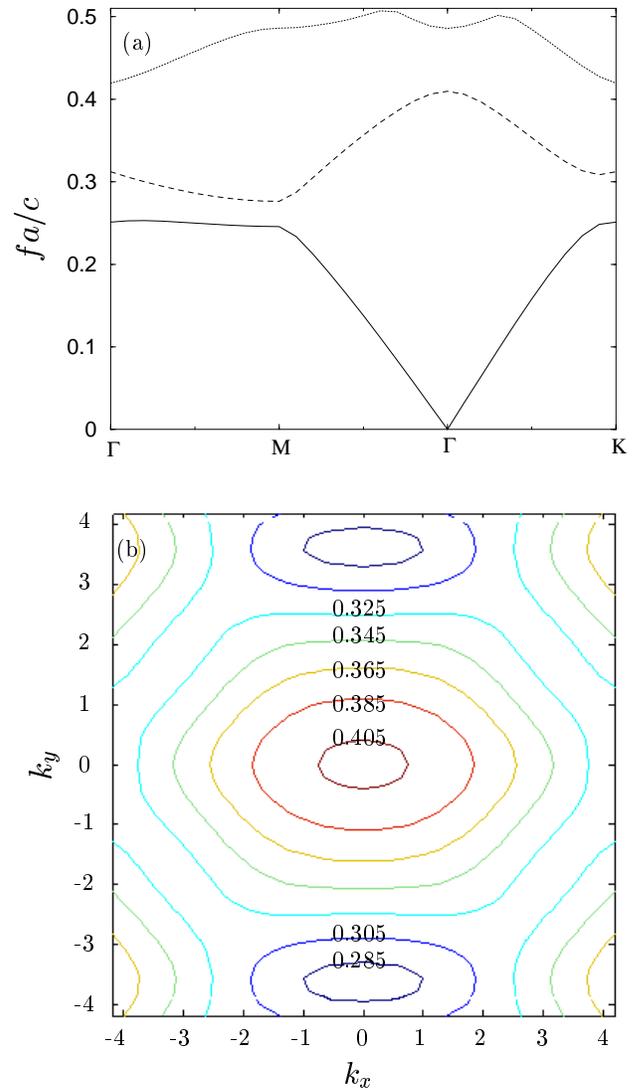}

\caption{( color online) (a) The photonic band structure for the TM polarization. (b) The
Equal Frequency Surfaces (EFS) with different dimensionless frequencies
shown on the corresponding curves. }
\end{figure}

Transmission measurements are performed to verify and test the negative
refraction and superlensing. The experimental setup consists of an
HP 8510C network analyzer, a waveguide horn antenna as the transmitter,
and a monopole antenna as the receiver. The PC used for the negative
refraction experiment has 9 layers in the propagation direction ($\Gamma M$)~and
33 layers in the lateral direction ($\Gamma M$). The interfaces are along the $\Gamma K$
direction. In all the experimental and theoretical results, the electric
field is kept parallel to the bars. The operational frequency that
leads to a structure of approximatively $n=-1$ is 6.5~GHz which correspond to a dimensionless frequency of $fa/c=0.345$. For the negative refraction
the horn antenna is oriented such that it makes an angle $\theta$ with
the normal to the $\Gamma K$ interface. 

To examine the negative refraction, we first measure the transmission
along the first interface in the $\Gamma K$ direction without the PC. We repeat
the measurement but this time with the PC and calculate the transmission  in the
vicinity of the second interface. The results are plotted
in Fig.~3. It is shown clearly how the center of the outgoing Gaussian
beam is shifted toward the left-handed side of the center of the
incident beam. This is a clear indication of the negative refraction
occurring inside the structure. To get a further insight about the negative or positive shift, Fig.~4 explains in details in which cases we have positive or negative refraction. 
\begin{figure}
\includegraphics[%
  scale=0.9]{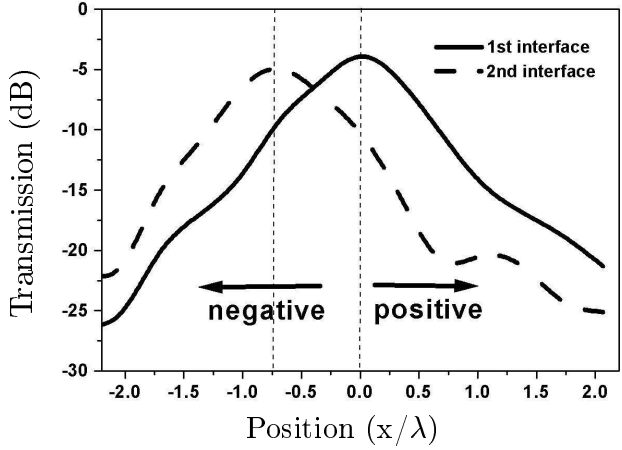}

\caption{The transmission at the first ( solid line) and second interface
(  dashed line) of the PC versus the lateral position. Arrows indicate
the refracted beam's direction depending on positive or negative shift.}
\end{figure}

It is important to notice that experimentally we did observe the negative refraction for different angles. This confirms that the negative refraction relies entirely on the LH behavior of the system and not on the convex shape of the EFS  \cite{key-5,key-7}. For a particular angle of $\theta=30^{\circ}$, the lateral shift was 3.2~cm which correspond to a negative angle inside the LH PC of  $-15^{\circ}$.

\begin{figure}
\includegraphics[%
  scale=0.7, clip]{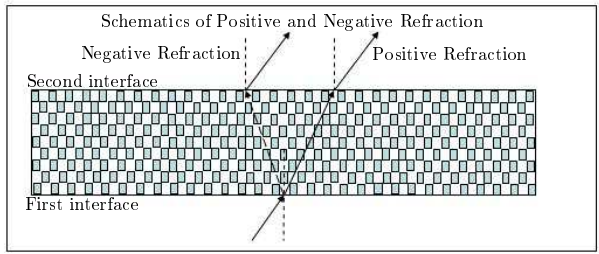}
\caption{( color online) Schematics of positive and negative refraction. For negative refracted beam, the center is shifted to the left hand side of the incident beam  and for the positive refracted beam, the center is shifted to the right hand side of the incident beam.}
\end{figure}

Further insight about the field distribution inside this structure
can be gained by observing the E field inside a LH PC's slab plotted in Fig.~5 after $77.33T$, where $T=2\pi/\omega$ is the period of the incident waves. The structure length along the lateral direction is about $13.8\lambda$
and about  $2.98\lambda$ in the propagation direction.
A Gaussian beam source with $\lambda=46~mm$ is placed outside the
PC making an angle $\theta=30^{\circ}$ with respect to the normal to the interface which is the $\Gamma M$ direction.
The FDTD simulation is performed by discretizing the real space into
a fine rectangular grid of ($a/54$ and $a/92$ for the x and y axes
respectively). A time step of $\delta t=24.90~ps$ is used. The distribution
of the field inside the PC shows how the beam is negatively refracted
inside the PC and positively refracted outside the PC parallel to
the incident beam. The theoretical result is in a good agreement with the experiment since it gives a negative refracted angle of about   $-15^{\circ}$.

\begin{figure}
\includegraphics[%
  scale=0.55]{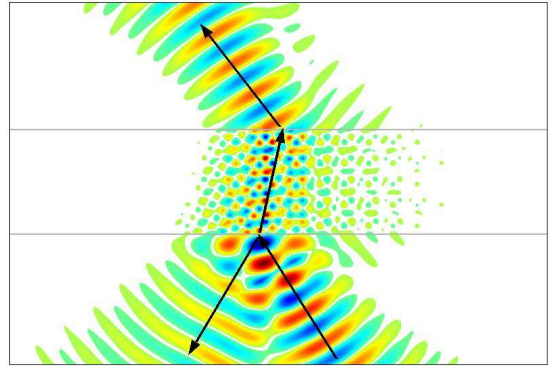}
\caption{ (color online) The E field of a Gaussian beam propagating along $30^{\circ}$ direction
after $77.33T$. The structure is $13.8\lambda$ in the lateral direction
and $2.98\lambda$ in the propagation direction. }
\end{figure}
An important advantage of this structure is its property to allow
single beam propagation. The structure was designed in such a manner
that only one beam propagation is allowed eliminating the undesired
Bragg reflections inside the PC\cite{key-5}. One of the interesting features of
this structure compared to the other LHM based on proposed PCs is
its high transmission. Fig.~6 displays the transmission versus the
dimensionless frequency. The experimental curve shows the high transmission
of the structure over a wide range of frequencies including the operational
one. The arrows in this figure show the theoretical indication
of the gaps. As predicted by the band structure, the experimental
curve shows two gaps in the $\Gamma M$ direction one between 0.25
and 0.276 and the other one starting at 0.41. Thus, the agreement
between the band structure and the experimental curve is very good.

\begin{figure}
\includegraphics[%
  scale=0.9]{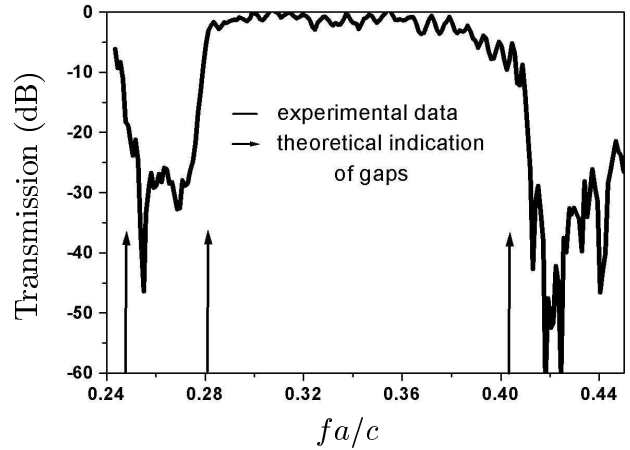}

\caption{ The transmission along $\Gamma M$ direction versus the dimensionless frequency. The solid arrows
indicate the theoretical gaps as calculated from the band structure.}
\end{figure}
In order to get an idea about what is happening inside the structure we examine the transmission in both directions. Fig.~7 shows the transmission in the $\Gamma M$ and $\Gamma K$ direction versus the frequency. This structure at the dimensionless operational frequency $(0.345)$ has a good transmission in the   $\Gamma M$ direction and a weak transmission in the  $\Gamma K$ direction. A weak transmission in the $\Gamma K$ direction enhance the transmission in the propagation direction which will be a critical issue for the superlensing phenomenon. It was argued \cite{key-14,key-15} that the focusing seen at the edge of the first band in a square lattice \cite{key-7} was not due to negative refraction but to anisotropic propagation resulting to the funneling effect. This is not the case in our  structure. As we will show latter, the focusing and the superlensing in our structure relies on negative refraction and on coupling to the surface states.\\  
\begin{figure}
\includegraphics[%
  scale=0.9]{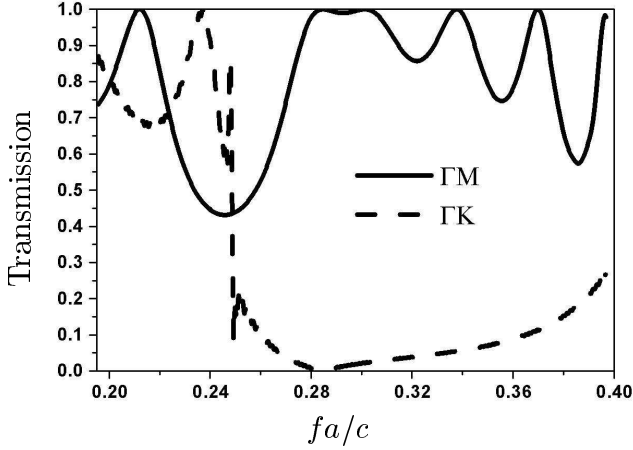}

\caption{The transmission versus the frequency for both $\Gamma M$ and $\Gamma K$ direction.}
\end{figure}
In order to investigate the superlensing phenomenon,
 we experimentally measure the intensity of the focus for the same
structure used in the transmission measurements ( 33 layers in the lateral direction and 9 layers
in the propagation direction). Two probes are used. The first one
is placed at a distance of $0.2\lambda$ from the first interface of the PC and
the second one at the same distance from the second interface. We
first measure the transmission with and without the PC . Fig.~8 shows
the two resulting curves. This figure demonstrates the high intensity
of the focus and that the transmission is enhanced by more than 20 dB for the setup with the PC compared to the one without the PC. The solid curve also shows
a high transmission at the edge of the structure at $x/\lambda\simeq1.75$. This might due to the
limited lateral length of the structure and to the support station which might give  some reflection at the edge. 

\begin{figure}
\includegraphics[%
  scale=0.9]{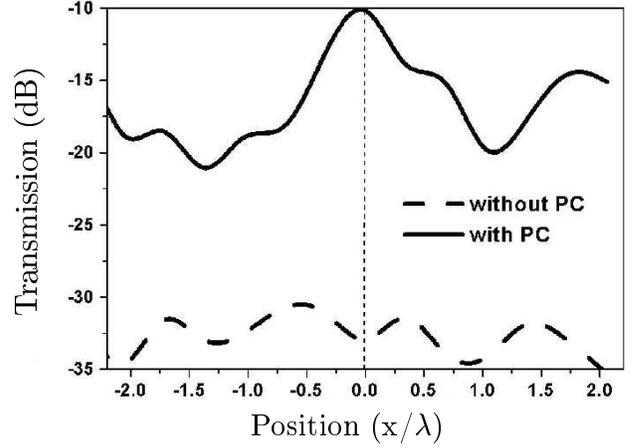}

\caption{The transmission versus the lateral position of a probe placed at a distance 
$0.2\lambda$ from the first interface without the PC (dashed curve
) and with the PC (solid curve) at 6.5~GHz or a dimensionless frequency $(fa/c)=0.345$.}
\end{figure}

Theoretically, we investigate the superlensing within a slab of PC
with a dimension of  $20.7\lambda$ in the lateral direction and $1.20\lambda$ in
the propagation direction. A point source with a dimensionless frequency $0.345$
placed at a distance  $0.15\lambda$ from the first interface of the PC is used.
A time step of $\delta t=26.5~ps$ and a fine rectangular discretization
mesh of $a/40$ and $a/92$ in the x and y direction are used. Fig.~9(a)
and Fig.~9(b) show the snapshots of the E field after $68.49 T$ for the complete
structure and the one with $0.10a$ termination. By the termination or cut $0.10a$ we mean that we cut  $0.10a$ from the first and last rows of the bars. Notice that in order to
excite the surface waves at the interface, we studied different
terminations of the surface \cite{key-16,key-17}. Among the different cuts, one surface
termination succeeds in exciting the surface waves as shown in Fig.~9b.
In this case the cut was about $0.10a$ . For both Figures ( Fig.~9a and 9b) a focus is obtained. However, a better image with
a higher intensity is obtained when  surface waves are involved.
The surface waves in Fig.~9b are clearly shown as propagating along
both the $\Gamma K$ interfaces. 

\begin{figure}
\includegraphics[%
  scale=0.7]{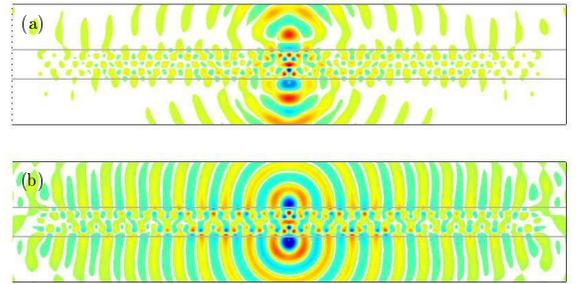}

\caption{(color online) Snapshots of the E field after $68.49T$ of a point source located
at $0.15\lambda$ and its image through a PC slab of $20.7\lambda$
in the lateral direction and $1.20\lambda$ in the propagation direction.
(a) shows the complete structure and (b) shows the structure with a surface cut of $0.10a$.}
\end{figure}
One way to verify the good quality of the image compared to the source
is to plot the average field intensity over a period at both the source
and image planes. Fig.~10 displays the normalized average field intensity
versus the lateral direction for the cut structure. The full width at half maximum (FWHM) of the image
beam is $0.35\lambda$ and it is almost the same as the FWHM at the source. This result shows that the cut structure, which supports surface waves focuses in perfect way the object. Thus with a
source that is not too close to the structure ($0.15\lambda$)we succeed in exciting
the evanescent waves and achieving a better focus.

\begin{figure}
\includegraphics[%
  scale=0.4,
  angle=-90]{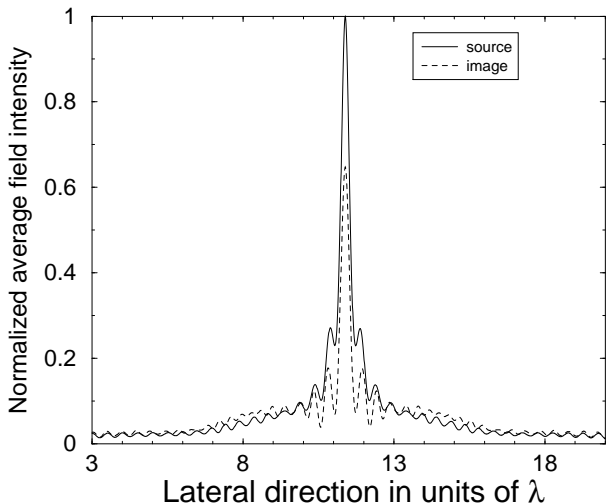}

\caption{The normalized average field intensity at the source (black curve)
and image plane (dashed curve) versus the lateral direction.}
\end{figure}

Apparently our structure with  $0.10a$  cut surface seems to be a good candidate for superlensing because we achieved a quite high transmission over a wide range of angles. Fig.~11 presents the transmission versus the angle of incidence for different values of surface terminations. By trying different terminations we managed to optimize the transmission. For the complete structure, the transmission start at 80\% for normal incidence, it gets high for a short range of angles then it drops to less than 20\% for angles larger than  $70^{\circ}$. However the structure with a surface termination of  $0.10a$ (see Fig.~11) shows the maximum transmission for all incident angles. Thus, the transmission is about 95\% for angles up to $40^{\circ}$, and more than 80\% for angles up to $65^{\circ}$ and more than 50\% for angles up to  $78^{\circ}$. This result shows first how the transmission gets enhanced by the means of the excitation of the surface waves for a non homogeneous structure. Second, high transmission over a wide range of incident angles excludes the possibility of the funneling effect \cite{key-14,key-15}and accentuate the fact that the negative refraction in our structure is purely a result of negative behavior \cite{key-5} and not a prefered propagation direction.
\begin{figure}
\includegraphics[%
  scale=0.9,
  ]{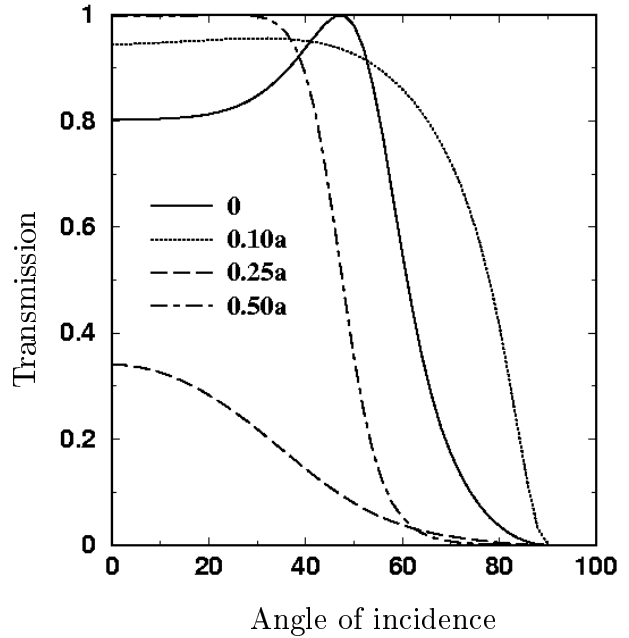}
\caption{Transmission versus the angle of incidence for different surface terminations.}
\end{figure}

In conclusion we have systematically studied a LH structure based
on a PC with a negative refractive index. Our experimental and
theoretical results show the negative refraction as well as the superlensing
phenomenon in this structure. Furthermore, surface waves at the
interface between air and the PC are excited within a specific termination
of the surface allowing the reconstruction of the evanescent waves
for better focus. The calculated average field intensity in the source
plane as well as in the image plane shows almost the same full width
at half maximum demonstrating the perfect image reproduced by this
new structure. Its high transmission along a wide range of angles
makes it a good candidate for observing LH behavior in PCs. 

This work was partially supported by Ames Laboratory (Contract No.
N-7405-Eng-82), DARPA (Contract No. MDA972-01-2-0016) and EU-DALHM.

\end{document}